\documentclass[reviewcopy]{elsart}

\usepackage{amsmath}
\usepackage{amsfonts}
\usepackage{amssymb}
\usepackage [ansinew] {inputenc}
\usepackage{graphicx}

\begin{document}

\begin{frontmatter}
\title{Substitution Systems \\ and Nonextensive Statistics}
\author{V.~Garc\'{\i}a-Morales\corauthref{cor1}}
\ead{garmovla@uv.es}
\corauth[cor1]{Corresponding author. Tel: +49 15224020629}
\address{Institute for Advanced Study - Technische Universit\"{a}t M\"{u}nchen, \\
Lichtenbergstr. 2a, D-85748 Garching, Germany}
\address{Departament de Termodin\`amica, Universitat de Val\`encia, \\ E-46100 Burjassot, Spain}
\begin{abstract}
\noindent{Substitution systems evolve in time by generating sequences of symbols from a finite alphabet: At a certain iteration step, the existing symbols are systematically replaced by blocks of $N_{k}$ symbols also within the alphabet (with $N_{k}$, a natural number, being the length of the $k$-th block of the substitution). The dynamics of these systems leads naturally to fractals and self-similarity. By using $\mathcal{B}$-calculus [V. Garcia-Morales, Phys. Lett. A 376 (2012) 2645] universal maps for deterministic substitution systems both of constant and non-constant length, are formulated in 1D. It is then shown how these systems can be put in direct correspondence with Tsallis entropy. A `Second Law of Thermodynamics' is also proved for these systems in the asymptotic limit of large words.}
\end{abstract}
\end{frontmatter}

\section{Introduction}

Substitution dynamical systems \cite{Queffelec,Fogg} are of great relevance to many branches of Physics and Mathematics including geometry, combinatorics, chaos and ergodic theory, spectral analysis, number theory, fractals and tilings.  The history of substitution systems dates back to 1906 with the following construction by A. Thue \cite{Thue} 
\begin{equation}
a \to ab \to abba \to abbabaab \to abbabaabbaababba \to \ldots  \label{Thue}
\end{equation}
This sequence, later rediscovered by M. Morse \cite{Morse} and therefore called the Thue-Morse sequence, can be easily obtained by the following iterative process: at each iteration replace each previous $a$ in the sequence by $ab$ and each $b$ by $ba$. This is probably  one of the simplest examples of a substitution system. 

\begin{figure}
\begin{center}
\includegraphics[width=0.4\textwidth, angle=270]{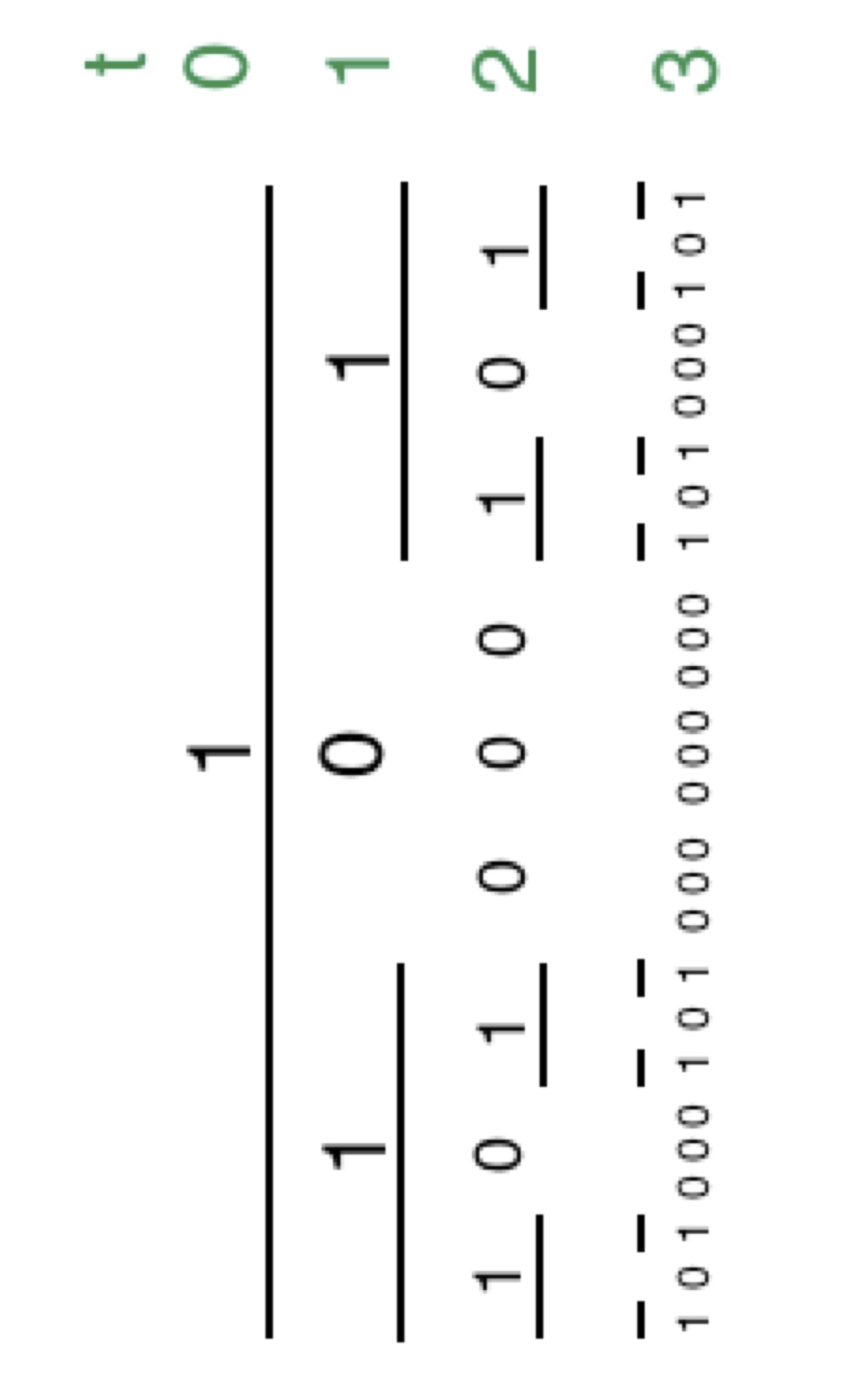}
\end{center}
\caption{\scriptsize{Construction of the Cantor set through a substitution process. At each iteration step $t$ the original segment is replaced by three segments that are one third shorter each, and the central one is removed. If the segments that remain at each iteration step are labelled by `1' and the segments that are removed are labelled by 0', this process has an associated binary sequence obtained by a simple substitution system.}} \label{Cantor}
\end{figure}

Another well-known example is suggested by the familiar construction leading to the Cantor set shown in Fig. \ref{Cantor}. This construction is related to a sequence of words containing only `0's and `1's obtained from a substitution system: The replacements that are made at each iteration step $t$ are $0 \to 000$ and $1 \to 101$. At $t=0$ one starts with `1' and at successive iteration steps one obtains the sequence of words $101$, $101000101$, $101000101000000000101000101$ and so on. Each of these words can thus be mapped to an odd natural number written in the binary base.

Both substitution systems are called of \emph{constant length} $N$, because the blocks of symbols used to replace the previous symbols have all the same size ($N=2$ in the Thue-Morse sequence and $N=3$ in the Cantor sequence). Thus, after $t$ iteration steps, the sequence contains $N^{t}$ symbols. Entire monographs have been dedicated to substitution systems of constant length (see for example \cite{Queffelec}). They are closely related to geometric tiling substitutions \cite{Natalie1,Mozes}, which are used to construct infinite tilings with a finite number of tile types. Since a single substitution rule is used in every iteration, substitution systems are also naturally related to fractals  \cite{Wolfram,Mandelbrot,Hutchinson,Barnsley,PeitgenBOOK}: If each symbol denotes a segment of, say, a given color, and the segment is systematically replaced by $N$ smaller segments that cover the previous one, the figure that will emerge as $t$ grows is expected to exhibit self-similarity, as the example of the Cantor sequence shows. 

Whereas there exist an extensive literature of constant-length substitution systems,
the ones with non-constant length have also a major importance and are receiving increased interest \cite{Natalie1}. The former are related to \emph{geometric} substitution tilings and the latter to \emph{combinatorial} substitution tilings \cite{Natalie1}. To give a simple illustration of a substitution system with non-constant length let us, for example, consider the following substitution rule of the symbols $a$ and $b$. At each step, one introduces the replacements  $a \to ab$ and $b \to a$. Then, if we begin with $a$ we get \cite{Natalie1}
\begin{equation}
a \to ab \to aba \to abaab \to abaababa \to abaababaabaab \to \ldots \label{sequnon}
\end{equation}
The length of the words is 1, 2, 3, 5, 8, 13, \ldots, i.e. the Fibonacci sequence. This length is no longer simply equal to $N^{t}$ at step $t$ as is the case in constant-length substitution systems but is, generally, more complicated. Still, a general expression for these systems can be found.

The outline of this article is as follows. In Section \ref{constant} we construct a universal map for constant-length substitution systems. In Section \ref{Tsallis} we show how these systems are naturally linked to entropy. Specifically, we show how Boltzmann and Tsallis entropies quite naturally arise from the dynamical evolution. Then, in Section \ref{nonco} we tackle the nontrivial problem of substitution systems with non-constant length and the article finishes with some considerations on the thermostatistics of such systems.

\section{$\mathcal{B}$-calculus and a universal map for substitution systems of constant length} \label{constant}

The main idea behind $\mathcal{B}$-calculus is to use a most elementary mathematical structure involving sums and/or products of so-called $\mathcal{B}$-functions as the basic building block to model computational processes \cite{VGM1}. This view proves quite useful to describe any rule-based dynamical system, as cellular automata \cite{VGM1,VGM2,VGM3}.   
 The $\mathcal{B}$-function for any real numbers $x$, $y$ is defined as \cite{VGM1}
\begin{equation}
\mathcal{B}(x,y) \equiv \frac{1}{2}\left(\frac{x+y}{|x+y|}-\frac{x-y}{|x-y|}\right) = \frac{1}{2}\left(\text{sign}(x+y)-\text{sign}(x-y)\right)  \label{d1}
\end{equation} 
with $\text{sign}(x) \equiv \frac{x}{|x|}$ being the sign function. Since $\text{sign}(0)\equiv 0$, at $x=\pm y$  (singular borders), $\mathcal{B}(\pm y,y)= \frac{y}{2|y|}= \text{sign}(y)/2$ and at the origin $\mathcal{B}(0,0) \equiv 0$. 


Let $S$ be the set of integers in the interval between $0$ and $p-1$, with $p > 1 \in \mathbb{N}$ denoting the alphabet size. We now consider an operator which replaces each possible value of a quantity $x \in S$ by a block of $N$ values $y_{0}$, $y_{1}$, ..., $y_{N-1}$ all in $S$. Thus, such an operator maps $S$ to $S^{N}$. Hence, if $h$ runs from $0$ to $N-1$, the substitution operator is defined as
\begin{equation}
y_{h}=\text{Subs}_{N;p}(x,h) \equiv \sum_{n=0}^{p-1}\sum_{m=0}^{N-1}a_{m+nN}\mathcal{B}\left(h-m,\frac{1}{2} \right)\mathcal{B}\left(n-x,\frac{1}{2} \right)=a_{h+Nx} \label{subope}
\end{equation}
where all $a_{m+nN} \in S$. The operator is uniquely specified by a Wolfram code $\text{Subs}_{N;p}$ with
\begin{equation}
\text{Subs} \equiv \sum_{n=0}^{p-1}\sum_{m=0}^{N-1}a_{m+nN}p^{m+nN}
\end{equation}
being a non-negative integer.

A substitution system is a substitution operator being iterated on a lattice that is also iteratively refined to accommodate each new output block: Each position on the lattice is labelled by a nonnegative integer $j \in [0, N_{t+1}-1]$ and has a dynamical state (a symbol of the alphabet) $u_{t}^{j} \in S$ at time $t$. Then, it gives rise to a block $_{N}\Sigma^{j}$ of $N$ positions at the next `time' step $t+1$. If the system has unit length at time $t = 0$ after $t$ time steps, since $N_{t}$ blocks have appeared, their thickness is reduced to $1/N_{t}$. A substitution system, as defined by Eq. (\ref{subope}), is therefore given by
\begin{equation}
u_{t+1}^{h+Nj}=\sum_{n=0}^{p-1}\sum_{m=0}^{N-1}a_{m+nN}\mathcal{B}\left(h-m,\frac{1}{2} \right)\mathcal{B}\left(n-u_{t}^{j},\frac{1}{2} \right) \label{subs}
\end{equation}
Then, from the dynamics of the substitution system, at time $t+1$ we have the replacement
$u_{t}^{j} \to u_{t+1}^{Nj}u_{t+1}^{1+Nj+1}\ldots u_{t+1}^{h+Nj+h}\ldots u_{t+1}^{(N+1)j-1}$
with every $u^{h}_{t+1}\in S$ integer. The label $h \in [0,N-1]$ marks the `position' of the symbol within the block. Since each digit is replaced by a block of $N$ new digits on the next iteration, now label $j$ gives the position of each new digit from the leftmost digit at $t+1$.

We now give three examples of the above map. The Thue-Morse sequence has $N = 2$ and $p = 2$ and, hence, it belongs to the family of $p^{pN}= 2^{4} = 16$ substitution systems $\text{Subs}_{2;2}$ with two symbols in the alphabet $p=2$ and block size $N =2$. We have $pN-1=3$ and $\textbf{a} = (a_{0}, a_{1}, a_{2}, a_{3}) = (0, 1, 1, 0)$ gives the rule. Thus $\text{Subs} = 6$ and the Thue-Morse sequence has code $6_{2;2}$.

The Cantor sequence can be obtained from the universal map Eq. (\ref{subs}) in a similar manner as the previous example for the Thue-Morse sequence. We have now $N = 3$ and $p = 2$. Since the Cantor sequence is given by the substitutions $0 \to 000$ and $1 \to 101$ this implies that $\textbf{a} = (a_{0}, a_{1},\ldots, a_{5}) =(0, 0, 0, 1, 0, 1)$ and thus $\text{Subs}=\sum_{n=0}^{p-1}\sum_{m=0}^{N-1}a_{m+nN}p^{m+nN}=\sum_{n=0}^{1}\sum_{m=0}^{2}a_{m+nN}2^{m+3n}=40$. Thus, the Cantor sequence in Fig. (\ref{Cantor}) is obtained from Eq. (\ref{subs}) for the substitution rule $40_{3;2}$ starting at $t = 0$ with just one symbol with value `1'.

\begin{figure}
\begin{center}
\includegraphics[width=0.25\textwidth, angle=270]{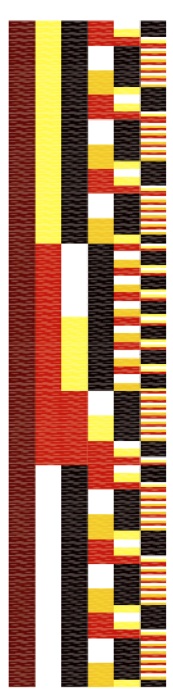}
\end{center}
\caption{\scriptsize{Evolution of substitution system $74330023345_{3;7}$ calculated from Eq. (\ref{subs}). Shown are $5$ iterations. Time flows from top to bottom}} \label{3s7}
\end{figure}

In this article we shall adhere to the following convention: If a number, say 13, is represented in any other radix $p$ that is not the decimal radix $p \ne 10$, then it shall be notated $(13)_{p}$.

The universal map, Eq. (\ref{subs}) can be used to derive \emph{any} substitution sequence with constant length in one dimension. We just only need the code of the rule to iterate Eq. (\ref{subs}). Let us consider, for example a rule with code $74330023345_{3;7}$. For $N = 3$ and $p = 7$ there is a huge total number of $7^{21}$ substitution rules. Since $pN = 21$, the vector $\mathbf{a}$ has 21 entries which correspond to the digits of the number $(5240652526264)_{7}$ (which equals $74330023345$ when written in radix $7$) in inverse order. Thus, we have $\textbf{a} = (a_{0}, a_{1}, ..., a_{21}) = (4, 6, 2, 6, 2, 5, 2, 5, 6, 0, 4, 2, 5, 0, 0, 0, 0, 0, 0, 0, 0)$. This encodes the following table of substitutions $0 \to 462$, $1 \to 625$, $2 \to 256$, $3 \to 042$, $4 \to 500$, $5 \to 000$ and $6 \to 000$. So, starting with one symbol with value `1', Eq. (\ref{subs}) provides the following sequence of words: $1 \to 625 \to 000256000 \to 462462462256000462462462$, etc. In Fig. \ref{3s7} the evolution of this substitution rule is represented. Five iterations of the rule are displayed from top to bottom. The color code is: black (`0'), brown (`1'), red (`2'), orange (`4') yellow (`5') and white (`6'). Symbol `3' never arises in the evolution of this rule: it can only be contained in the initial condition.

\section{Connection to Tsallis nonextensive thermostatistics} \label{Tsallis}

Substitution systems provide examples where Tsallis nonextensive statistics \cite{Tsallis,Tsallis2,VGMStat,Olemskoi} can be
naturally applied. In \cite{VGMStat} we found that when a stationary state is reached ($t \to \infty$) fractal phase spaces of box-counting dimension $D$ lead naturally to the Tsallis' entropic form $S_{q}$ \cite{Tsallis} in the microcanonical ensemble, 
\begin{equation}
S_{q}=\frac{W^{1-q}-1}{1-q} \label{superTsallis}
\end{equation}
with $W$ being the total number of microstates, $q = D/D^{*}$ being the entropic parameter \cite{VGMStat} and $D^*$ being the dimension of the phase space where the fractal object is embedded. 

The entropic parameter can be explicitly calculated for 1D substitution systems. Let us consider substitution rules $\text{Sub}_{2;N}$ acting on two symbols `0' or `1'. These symbols may denote whether a specific microstate is actually reached or not through the dynamical trajectory of the system, that we assume to be a fractal object, as described in \cite{VGMStat}. The box-counting dimension of the pattern $D$ that arises from the substitution rule is calculated by counting the total number $\mathcal{N}_{1}$ of positions $j$ where a symbol `1' is located. Since at time $t$ there is a total number $W=N_{t}$ of boxes, we have, from the definition of box-counting dimension
\begin{equation}
D\equiv \lim_{t\to \infty} \frac{\ln \mathcal{N}_{1}}{\ln W} =\lim_{t\to \infty} \frac{\ln \mathcal{N}_{1}}{\ln N^{t}} \label{pbc}
\end{equation}
Now, from Eq. (\ref{subs}), starting from an initial condition $u_{0}^{0} = 1$ we have, at time 
$t$
\begin{eqnarray}
\mathcal{N}_{1}&=&\sum_{j=0}^{N^{t-1}-1}\sum_{h=0}^{N-1}u_{t}^{Nj+h}\mathcal{B}\left(u_{t}^{Nj+h}-1,\frac{1}{2}\right) \nonumber \\
&=&\sum_{j=0}^{N^{t-1}-1}\sum_{h=0}^{N-1}\sum_{m=0}^{N-1}a_{m+N}\mathcal{B}\left(h-m, \frac{1}{2} \right)  \mathcal{B}\left(u_{t-1}^{j}-1, \frac{1}{2} \right)\mathcal{B}\left(u_{t}^{Nj+h}-1,\frac{1}{2}\right) \nonumber \\
&=& \left(\sum_{m=0}^{N-1}a_{m+N}\right)^{t} \label{prue}
\end{eqnarray}
where we have used that a value `1' cannot have a value `0' as predecessor. Now, by using Eqs. (\ref{pbc}) and (\ref{prue}) we obtain
\begin{equation}
D =\lim_{t\to \infty} \frac{\ln \left(\sum_{m=0}^{N-1}a_{m+N}\right)^{t}}{\ln N^{t}}=\frac{\ln\left(\sum_{m=0}^{N-1}a_{m+N}\right)}{\ln N} =q \label{bc}
\end{equation}
Thus, since $D^{*} = 1$ in this case, we have an explicit expression for the entropic parameter $q$. For example, for the Cantor substitution rule $40_{2;3}$, since the rule has vector $a = (a_{0}, a_{1},\ldots , a_{5}) =(0, 0, 0, 1, 0, 1)$ we obtain, from Eq. (\ref{bc})
\begin{equation}
D=\frac{\ln\left(\sum_{m=0}^{N-1}a_{m+N}\right)}{\ln N}=\frac{\ln\left(a_{3}+a_{5}\right)}{\ln 3}=\frac{\ln 2}{\ln 3}=0.6309  
\end{equation}
which is the well known fractal dimension of the Cantor set.

We thus observe that substitution systems are naturally expected to obey nonextensive statistics \cite{Tsallis,Tsallis2,VGMStat,Olemskoi}. Compared to featureless and uncorrelated dynamical systems described by Boltzmann entropy, substitution systems contain more information through the self-similarity property, which translates to all scales once the stationary state has been reached. This property is responsible for the value $q \in [0,1]$ which qualitatively describes the correlations in phase space, encoding the self-similarity property of the fractal occupancy of the available microstates \cite{VGMStat, VGMmacroion}. Such correlations have been described in e.g. electrochemical systems, when describing the distribution of counterions in an electrolyte close to a charged planar interface \cite{VGMmacroion} or in the non-Markovian charge transfer processes to metallic nanoelectrodes \cite{PNAS2}.

\section{Universal map for non-constant-length substitution systems and their thermodynamic limit} \label{nonco}

Substitution systems of non-constant length are also quite important for Physics (for example for the mathematical description of quasicrystals \cite{quasi}). These systems replace each symbol in the alphabet by blocks of symbols with varying thicknesses so that, the block thickness $N$ is no longer a constant but a function $N_{n}$ of the symbol value $n \in [0,p-1]$. Their resulting evolution is far more complex than the observed for constant-length ones. However, the results presented in Section \ref{constant} can be fruitfully exploited to address the more demanding problem of providing a universal map for these systems (from which the constant-length ones are a subset).

Since the blocks can vary in length, it is necessary to have a criterion for the map to stop the substitutions (the map should `know' where a word begins and ends). There are two possibilities to achieve this: 1) we can exclude the symbol `0' from the alphabet and consider only mappings of the symbols in the set $S^{*} \in [1,p-1]$; 2) we can design the substitution system so as to obey the following conditions
\begin{itemize}
\item The symbol 0 is mapped to itself $0 \to 0$ or to a block of zeros of length $N_{0}$.
\item Any nonzero symbol $n$ is mapped to a block of length $N_{n}$ were the leftmost digit (i.e. the most significant digit) is nonzero.
\end{itemize}
We consider in the following this second possibility. Clearly, the Cantor sequence obeys these two conditions since $0 \to 000$ and $1 \to 101$. The strategy that we shall now follow is indeed suggested by the Cantor sequence in Fig. \ref{Cantor}: We shall map words to natural numbers exploiting number theoretic properties of natural numbers. Thus, the total length of a word is the total number of digits of the corresponding natural number and equals the exponent of the power of the base accompanying the most significant digit plus one (see below). 

We now introduce some more concepts that we shall need. Let $A$ be a natural number,  
the digit function $\mathbf{d}_{p}(k,A)$, for $p \ge 1 \in \mathbb{N}$ and $k$ a non-negative integer is defined as \cite{QUANTUM}
\begin{equation}
\mathbf{d}_{p}(k,A)=\left \lfloor \frac{A}{p^{k}} \right \rfloor-p\left \lfloor \frac{A}{p^{k+1}} \right \rfloor    \label{cucuAreal}
\end{equation}
and gives the $k$-th digit of the natural number $A$ (when it is non-negative) in a positional numeral system in radix $p > 1$. If $p=1$ the digit function satisfies $\mathbf{d}_{1}(k,A)=\mathbf{d}_{1}(0,A)=0$ and it does not relate to a positional numeral system. In Eq. (\ref{cucuAreal}) $\lfloor \ldots \rfloor$ denotes the floor function (lower closest integer) of the quantity between the brackets. 

With the digit function we can express any natural number $A$ as \cite{QUANTUM} 
\begin{equation}
A=\sum_{k=0}^{\lfloor \log_{p}A \rfloor} p^{k} \mathbf{d}_{p}(k,A) \label{idenreal}
\end{equation}

\noindent \emph{Example:} In the decimal radix $p=10$, the number $A=8674$ has digits $\mathbf{d}_{10}(0,A)=4$, $\mathbf{d}_{10}(1,A)=7$, $\mathbf{d}_{10}(2,A)=6$, $\mathbf{d}_{10}(3,A)=8$.  

We can easily see that Eq. (\ref{idenreal}) holds since, by using Eq. (\ref{cucuAreal})
\begin{eqnarray}
&&\left \lfloor \frac{A}{p^{k}} \right \rfloor-p\left \lfloor \frac{A}{p^{k+1}} \right \rfloor =\left \lfloor \frac{\sum_{k'=0}^{\lfloor \log_{p}A \rfloor} p^{k'} \mathbf{d}_{p}(k',A)}{p^{k}} \right \rfloor-p\left \lfloor \frac{\sum_{k'=0}^{\lfloor \log_{p}A \rfloor} p^{k'} \mathbf{d}_{p}(k',A)}{p^{k+1}} \right \rfloor \nonumber \\
&=& \left \lfloor \sum_{k'=0}^{\lfloor \log_{p}A \rfloor} p^{k'-k} \mathbf{d}_{p}(k',A) \right \rfloor-p\left \lfloor \sum_{k'=0}^{\lfloor \log_{p}A \rfloor} p^{k'-k-1} \mathbf{d}_{p}(k',A) \right \rfloor \nonumber \\
&=& \sum_{k'=k}^{\lfloor \log_{p}A \rfloor} p^{k'-k} \mathbf{d}_{p}(k',A) -p \sum_{k'=k+1}^{\lfloor \log_{p}A \rfloor} p^{k'-k-1} \mathbf{d}_{p}(k',A)=\mathbf{d}_{p}(k,A)  \nonumber 
\end{eqnarray}
The upper bound in the sum Eq. (\ref{idenreal}) gives the total number of integer digits of $A$ when written in radix $p>1$ since it is equal to $1+\lfloor \log_{p}A \rfloor$. It is easy to prove this since for $p > 1$, we have $\lfloor \log_{p}A \rfloor \le  \log_{p}A  < \lfloor \log_{p}A \rfloor+1$. Thus, $p^{\lfloor \log_{p}A \rfloor} \le  A  < p^{\lfloor \log_{p}A \rfloor+1}$ and, therefore, $\left \lfloor \frac{A}{p^{k}} \right \rfloor=0 \qquad \forall k > \lfloor \log_{p}A \rfloor$. From the definition this in turn implies that most significant exponent of $p$ contributing to the sum is given by $\lfloor \log_{p}A \rfloor$. Thus,
\begin{equation}
 \mathbf{d}_p(k,A) = 0 \quad \qquad \forall k > \lfloor \log_{p}A \rfloor  \label{bound}  
\end{equation}
which indicates that the total number of digits is equal to $1+\lfloor \log_{p}A \rfloor$.

A useful operator is the concatenation operator $J_{p}(A,B)$. This is a 2-variable operator acting on the non-negative integers $A$ and $B$ given in radix $p$ to produce another non-negative integer, which contains the digits of $B$ appended after the most significant digit of $A$. This operator is thus defined as
\begin{equation}
J_{p}(A,B)\equiv Bp^{1+\lfloor \log_{p}A \rfloor}+A \label{juxtop}
\end{equation}
As example we have $J_{10}(2149,\ 987605)=9876052149$. As a mathematical curiosity, from this definition one easily checks that
\begin{equation}
\log_{p}\left(\frac{J_{p}(A,B)-A}{B}\right)=1+\lfloor \log_{p}A \rfloor
\end{equation}
is always a natural number (that gives the number of digits of $A$ written in radix $p$). One has $J_{p}(A,B)=J_{p}(B,A)$ only if $A=B$, in which case this latter result becomes
\begin{equation}
\log_{p}\left(\frac{J_{p}(B,B)}{B}-1\right)=1+\lfloor \log_{p}B \rfloor
\end{equation}
\noindent \emph{Example:} $\log_{10}\left(\frac{8546744385467443}{85467443}-1\right)=8$ (i.e. the number of digits of $85467443$ in the decimal radix).


Let now $A_{t}$ denote a natural number representing the whole word obtained through the substitution system at time $t$. We have $u_{t}^{j}= \mathbf{d}_p(j,A_{t})$ where $u_{t}^{j} \in S$ has the same meaning as in Section \ref{constant}. This symbol is replaced by a word represented by the natural number 
\begin{eqnarray}
f(u_{t}^{j})&=&\sum_{k=0}^{p-1}\sum_{h=0}^{N_{k}-1}p^{h}\text{Subs}_{N_{k}}(u_{t}^{j}, h)\mathcal{B}\left(k-u_{t}^{j}, \frac{1}{2}\right) \nonumber \\
&=&\sum_{k=0}^{p-1}\sum_{h=0}^{N_{k}-1}p^{h}\text{Subs}_{N_{k}}(\mathbf{d}_p(j,A_{t}), h)\mathcal{B}\left(k-\mathbf{d}_p(j,A_{t}), \frac{1}{2}\right)=f(\mathbf{d}_p(j,A_{t})) \label{lac}
\end{eqnarray}
where $\text{Subs}_{N_{k}}(u_{t}^{j}, h)$ is given by Eq. (\ref{subs}). Note the dependence of the length of the substitution on $k \in [0,p-1]$. 

The complete word at time $t+1$, $A_{t+1}$ is formed by concatenating all different $f(\mathbf{d}_p(j,A_{t}))$ from $j=0$ to $\lfloor \log_{p}A_{t} \rfloor$. This amounts to the composition of the concatenation operator with itself $\lfloor \log_{p}A_{t} \rfloor$ times. We thus finally obtain
\begin{equation}
A_{t+1}=\sum_{m=0}^{\lfloor \log_{p}A_{t} \rfloor}p^{m+\sum_{j=0}^{m-1}\lfloor \log_{p}f(\mathbf{d}_p(j,A_{t})) \rfloor}f(\mathbf{d}_p(m,A_{t})) \label{themap}
\end{equation}
for the map $A_{t} \to A_{t+1}$ between words. Eq. (\ref{themap}) with $f$ given by Eq. (\ref{lac}) is the map sought and the main result of this paper. This can be proved by induction as we show next. If the length of the word is one symbol, then $\lfloor \log_{p}A_{t} \rfloor=0$ and from the map above we have $A_{t+1}=f(\mathbf{d}_p(0,A_{t}))$ as it must be. Then let us assume the map valid for $n-1=\lfloor \log_{p}A_{t} \rfloor-1$ concatenations. For $n=\lfloor \log_{p}A_{t} \rfloor$ concatenations then, we find by using Eq. (\ref{juxtop})
\begin{eqnarray}
&&A_{t+1}=J_{p}\left(f(\mathbf{d}_p(n,A_{t})), \sum_{m=0}^{\lfloor \log_{p}A_{t} \rfloor-1}p^{m+\sum_{j=0}^{m-1}\lfloor \log_{p}f(\mathbf{d}_p(j,A_{t})) \rfloor}f(\mathbf{d}_p(m,A_{t})\right) \nonumber \\
&&=f(\mathbf{d}_p(n,A_{t}))p^{\lfloor \log_{p}A_{t} \rfloor+\sum_{j=0}^{m-1}\lfloor \log_{p}f(\mathbf{d}_p(j,A_{t}))} +\sum_{m=0}^{\lfloor \log_{p}A_{t} \rfloor-1}p^{m+\sum_{j=0}^{m-1}\lfloor \log_{p}f(\mathbf{d}_p(j,A_{t})) \rfloor}f(\mathbf{d}_p(m,A_{t}) \nonumber \\
&&=\sum_{m=0}^{\lfloor \log_{p}A_{t} \rfloor}p^{m+\sum_{j=0}^{m-1}\lfloor \log_{p}f(\mathbf{d}_p(j,A_{t})) \rfloor}f(\mathbf{d}_p(m,A_{t}))
\end{eqnarray}
which is Eq. (\ref{themap}), as we wanted to prove.

As an example, let us consider the substitution rule of non-constant length described in the introduction $a \to ab$, $b \to a$. We take $f(0)=0$ and denote $a$ by `1' and $b$ by `2'. Thus, the alphabet consists of the symbols 0, 1, 2 (although if we start with a nonzero initial condition the symbol '0' never arises in the sequence), having size $p=3$. Then, we have $f(1)=12$ and $f(2)=1$ and if we start from `1', then $A_{0}=1=\mathbf{d}_3(0,A_{0})$ and Eq. (\ref{themap}) produces the sequence of natural numbers
\begin{equation}
(1)_{3} \to (12)_{3} \to (121)_{3} \to (12112)_{3} \to (12112121)_{3} \to (1211212112112)_{3} \to \ldots
\end{equation}
which represent the words in the sequence of Eq. (\ref{sequnon}). Note that we can select any $p >3$ and Eq. (\ref{themap}) will still yield the same sequence (provided that one starts only with the symbols `1' and `2').

We note that if $N_{k}=N$ for all $k$ in Eq. (\ref{lac}), that equation then reduces to
\begin{eqnarray}
f(\mathbf{d}_p(j,A_{t}))&=&\sum_{h=0}^{N-1}p^{h}\text{Subs}_{N}(u_{t}^{j}, h)=\sum_{h=0}^{N-1}p^{h}\text{Subs}_{N}(\mathbf{d}_p(j,A_{t}), h)\label{lac2}
\end{eqnarray}
and, since $\lfloor \log_{p}f(\mathbf{d}_p(j,A_{t})) \rfloor=N-1$, we obtain from Eq. (\ref{themap})
\begin{equation}
A_{t+1}=\sum_{m=0}^{\lfloor \log_{p}A_{t} \rfloor}\sum_{h=0}^{N-1}p^{h+Nm}\text{Subs}_{N}(\mathbf{d}_p(m,A_{t}), h) \label{themap2}
\end{equation}
which clearly corresponds to the map governing the evolution of the words for a constant-length substitution. This can be seen by noting that, from this latter equation
\begin{equation}
u_{t+1}^{h+Nj}=\mathbf{d}_{p}(h+Nj,A_{t+1})=\text{Subs}_{N}(\mathbf{d}_p(j,A_{t}), h)=\text{Subs}_{N}(u_{t}^{j}, h) 
\end{equation}
which coincides with Eq. (\ref{subs}).

When $A_{t}$ is very large, since $f(\mathbf{d}_p(m,A_{t}))$ is bounded, $f(\mathbf{d}_p(m,A_{t})) <<A_{t}$ , Eq. (\ref{themap}) behaves asymptotically as
\begin{eqnarray}
A_{t+1}&\sim& p^{\lfloor \log_{p}A_{t} \rfloor+\sum_{j=0}^{\lfloor \log_{p}A_{t} \rfloor-1}\lfloor \log_{p}f(\mathbf{d}_p(j,A_{t})) \rfloor}f(\mathbf{d}_p(\lfloor \log_{p}A_{t} \rfloor,A_{t})) \nonumber \\
&\sim& p^{\sum_{j=0}^{\lfloor \log_{p}A_{t} \rfloor-1}\lfloor \log_{p}f(\mathbf{d}_p(j,A_{t})) \rfloor}A_{t}f(\mathbf{d}_p(\lfloor \log_{p}A_{t} \rfloor,A_{t})) 
\label{themapasin}
\end{eqnarray}   
and, therefore
\begin{equation}
\frac{A_{t+1}}{A_{t}} \sim p^{\sum_{j=0}^{\lfloor \log_{p}A_{t} \rfloor-1}\lfloor \log_{p}f(\mathbf{d}_p(j,A_{t})) \rfloor}f(\mathbf{d}_p(\lfloor \log_{p}A_{t} \rfloor,A_{t})) \label{asin}
\end{equation}
This expression quantifies the asymptotic growth in the length of the words of the substitution sequences. If we consider this equation as a model of natural physical systems we can now elucidate why irreversibility generally takes place by means of the principle of least radix economy. We have recently formulated this principle as a unified approach to classical and quantum physics \cite{QUANTUM}. The principle of least radix economy establishes that physical laws are those for which the physical radix given by $p=\eta=\left \lfloor \frac{S}{h} \right \rfloor$ (where $S$ is the Lagrangian action and $h$ Planck's constant) is most efficient. This efficiency or `economy' is quantified by the radix economy (also called digit capacity) \cite{QUANTUM,Hurst} which is defined as
\begin{equation}
\mathcal{C}\left(\eta, \mathcal{A} \right)= \eta \left \lfloor1+\log_{\eta} \mathcal{A} \right \rfloor  \label{capac}
\end{equation}
where $\mathcal{A}$ is a Lagrangian action functional (radix dependent) or, simply, just a number (radix independent) that counts all available configurations in phase space. 
The principle of least radix economy then establishes that physical laws are those which extremize the radix economy $\mathcal{C}\left(\eta, \mathcal{A} \right)$ above.
 
The main prediction of the principle of least radix economy relevant to us here is that asymptotically, when $\mathcal{A}$ is a large number representing the total number of microstates on the available constant energy surface (in our case $\mathcal{A}=A_{t}$) the physical radices with the least economy are, on the average, $\eta= \lfloor e \rfloor=2$ or $\eta= \lfloor e \rfloor +1=3$ (see Section 8 in \cite{QUANTUM}). This means that the optimal size of the alphabet in such asymptotic situation is either 2 or 3 in the average, which means that if we represent physical processes as substitution systems, we have long words containing only the symbols '0' and '1', or '0', '1' and '2'. \emph{This generally leads to either a constant or an increased size of the words in the substitution sequence}. To show this, let us consider an hypothetical situation where e.g. the substitutions
\begin{equation}
(12)_{p_{t}} \to (1)_{p_{t}} \qquad (1)_{p_{t}} \to (12)_{p_{t}}
\end{equation}
where both equally possible. Here the radix $p_{t}$ is allowed to depend on the discrete time step $t$ in a random manner. Taking this into account, the first of these substitutions would mean a contraction of the phase space. But for such a substitution to be possible, the physical radix $p_{t}=\eta$ would have to be $p_{t} \ge 13$: Note that consistency demands that what is \emph{to the left} of the arrow in a substitution rule is just only one symbol of the alphabet and so $(12)_{p_{t}}$ can only be a symbol of the alphabet if $p_{t} \ge 13$. Such a value for the radix is much larger than the most economical one $p_{t} \approx e$ in the asymptotic regime because of the principle of least radix economy (see \cite{QUANTUM}, Section 8). Thus, if one accepts that principle as a valid description of physical laws, transitions as $(12)_{p_{t}} \to (1)_{p_{t}}$ have a negligible measure in the available phase space in the asymptotic regime and only length-preserving or length-expanding words such as $(1)_{p_{t}} \to (12)_{p_{t}}$ are observed with significant probability (for any arbitrary substitution system, if it is to have a physical meaning). In brief, the principle of least radix economy implies that Eq. (\ref{asin}) takes, for a physical process, the asymptotic form
\begin{equation}
\frac{A_{t+1}}{A_{t}} \sim e^{\sum_{j=0}^{\lfloor \ln A_{t} \rfloor-1}\lfloor \ln f(\mathbf{d}_p(j,A_{t})) \rfloor}f(\mathbf{d}_p(\lfloor \ln A_{t} \rfloor,A_{t})) \ge 1 \label{asinb}
\end{equation}
with
\begin{equation}
\sum_{j=0}^{\lfloor \ln A_{t} \rfloor-1}\lfloor \ln f(\mathbf{d}_p(j,A_{t})) \rfloor \ge 0 \label{asinc}
\end{equation}
This proves a `Second Law of Thermodynamics' for any physically meaningful substitution system: Only if the substitution has constant length $N=1$ and the substitution is a bijection $S \to S$ can the behavior be reversible. In this case, we would have the equal sign in Eqs. (\ref{asinb}) and (\ref{asinc}).


\section{Conclusions}

In this note a universal map for all one-dimensional deterministic substitution systems has been derived by means of $\mathcal{B}$-calculus \cite{VGM1,VGM2,VGM3}: Our main result Eq. (\ref{themap}) describes both substitutions of constant and non-constant length. The thermodynamic limit of the map has also been studied: By means of the principle of least radix economy, recently introduced \cite{QUANTUM} and the methods proposed here, we have proved the Second Law of Thermodynamics for these systems. We have also shown how constant-length substitution systems are naturally connected to Tsallis nonextensive thermostatistics \cite{Tsallis} through the phase space occupancy of a fractal object. Our findings presented here give additional support to our previous approach to generalized thermostatistics \cite{VGMStat} providing also an explicit example where the essential discreteness that has been found in connection with Tsallis thermostatistics \cite{Abe} seems most natural.

Substitution systems as presented here can act on arbitrary topologies. In our proofs only general number theoretic properties have been invoked and, although we have used a representation involving digits of natural numbers by means of positional number systems in radix $p$, the actual spatial position of the quantities that the digits may represent in a physical system is immaterial for our approach. Thus, interaction of substitution systems with cellular automata, tessellations of the plane and dynamics on Voronoi diagrams are further directions in which the results presented in this article might be further expanded. It would also be interesting to analyze in detail the substitution maps here obtained in relationship with irreversibility and entropy production as applied also to problems in biophysics and chemical physics \cite{Roldan,Seifert3,Hoover,Hoover2,Annals,KanekoPHYSA,Bennett}. Finally, because of the connection with Tsallis statistics here established, it might be interesting to address the problem of how other forms of superstatistics \cite{Beck} and spectral statistics \cite{TsallisSPECTRAL} can arise out of the evolution of substitution systems.

\bibliography{biblos}{}
\bibliographystyle{h-physrev3.bst}

\end{document}